\documentclass[]{spie}
\usepackage[]{graphicx}
\usepackage{subfigure,amsmath, amssymb,amsfonts}
\usepackage{wrapfig}
\usepackage{wrapfig}
\usepackage{color}
\usepackage[dvipsnames]{xcolor}
\usepackage[colorlinks=true, allcolors=blue]{hyperref}
\title{The Exoplanet Transmission Spectroscopy Imager (ETSI)}

\author[a]{Mary Anne Limbach*\footnote{The two leading authors contributed equally to the optical design of ETSI}}
\author[a]{Luke M. Schmidt*}
\author[a]{D. L. DePoy}
\author[a]{Jeffrey C. Mason}
\author[b]{Mike Scobey}
\author[b]{Pat Brown}
\author[a]{Chelsea Taylor}
\author[a]{Jennifer L. Marshall}
\affil[a]{George P. and Cynthia Woods Mitchell Institute for Fundamental Physics and Astronomy, and Department of Physics and Astronomy, Texas A\&M University, College Station, TX 77843-4242, USA}
\affil[b]{Alluxa Inc, 3660 North Laughlin Rd, Santa Rosa, CA 95403 USA}

\begin{document}
\maketitle

\begin{abstract}
We present the design of a novel instrument tuned to detect transiting exoplanet atmospheres. The instrument, which we call the exoplanet transmission spectroscopy imager (ETSI), makes use of a new technique called common-path multi-band imaging (CMI). ETSI uses a prism and multi-band filter to simultaneously image 15 spectral bandpasses on two detectors from $430-975nm$ (with a average spectral resolution of  $R = \lambda/\Delta\lambda = 23$) during exoplanet transits of a bright star. A prototype of the instrument achieved photon-noise limited results which were below the atmospheric amplitude scintillation noise limit.  ETSI can detect the presence and composition of an exoplanet atmosphere in a relatively short time on a modest-size telescope. We show the optical design of the instrument. Further, we discuss design trades of the prism and multi-band filter which are driven by the science of the ETSI instrument. We describe the upcoming survey with ETSI that will measure dozens of exoplanet atmosphere spectra in $\sim2$ years on a two meter telescope. Finally, we will discuss how ETSI will be a powerful means for follow up on all gas giant exoplanets that transit bright stars, including a multitude of recently identified TESS (NASA's Transiting Exoplanet Survey Satellite) exoplanets.
\end{abstract}

\section{Introduction}
Characterization of exoplanets via spectral measurement is key to understanding their atmospheric composition, weather, variability/climate and potential habitability. Two methods, transit spectroscopy and direct imaging integral field spectroscopy (IFS), have proven successful techniques for characterizing exoplanet atmospheres. An exoplanet transmission spectra is measured during an exoplanet's transit. The exoplanet’s atmosphere absorbs some of the stellar light passing through the exoplanets atmosphere and imprints the exoplanet's absorption spectrum onto the stellar light\cite{Seager2000, Kreidberg2017}. Observations of exoplanet transit spectra with space-based observations has produced reliable results\cite{Deming2013, Kreidberg2014, Ingalls2016, Sing2016}. Ground based transmission spectroscopy measurements have proven more difficult with results often difficult to reproduce or disputed\cite{Tinetti2007, Swain2008, Gibson2011}. However, more recent ground-based exoplanet transit observations show excellent photometric precisions\cite{Kirk2017, Kirk2018}, but require a large ($\sim 8$ meter) telescope to perform observations. Ground-based observations are typically limited by many systematic noise sources with precisions often many times worse than the photon noise limit. The largest source of systematic noise typically arises from the time-variability of the earth's atmosphere\cite{Li2016}. 

In this paper we discuss the design of a instrument, the Exoplanet Transmission Spectroscopy Imager (ETSI). This instrument uses a novel method to perform exoplanet transmission spectroscopy aimed at both (1) eliminating the systematic noise due to the earth's atmosphere as well as all other sources of achromatic systematic noise and (2) operating on small to medium class telescopes ($D = 2-4$ meters) to increase the viability of surveying and cataloging a large numbers of exoplanet atmospheres. In Section \ref{CMI} we discuss the novel technique used to minimize systematic noise. In Section \ref{designETSI} we discuss the optical design of the ETSI instrument with a detailed examination of the ETSI multi-band filter and prism, which are key to ETSI's unique design, in Section \ref{FilterDesign}. Finally, in Section \ref{science} we discuss the expected science with ETSI including our observing plan and simulated ETSI exoplanet transit spectra results.

\section{The Common-Path Multi-band Imaging (CMI) Technique}\label{CMI}
ETSI will make use of a new technique for transit spectroscopy called common-path multi-band imaging (CMI) to characterize exoplanet atmospheres. The CMI method uses a multi-band (15 spectral bands) picket-fence filter and low-dispersion prism to image discrete, well separated point spread functions (PSFs) of a star in all 15 bands simultaneously on two detectors. The resulting image from this unique instrument is shown in Figure \ref{opticalLayout}, upper left. This is a simulated ETSI image at one of the two detectors, which images half of the ETSI spectral bands.

Because the spectral bands are well separated PSFs on the image, it is possible to measure the flux in each PSF separately and then compare the flux of one PSF to another {\it of the same star but in different spectral bands}. This can be done for all spectral bands of the star allowing for relative color measurements of the star in all 15 bands without the use of any reference star (which introduces systematic noise). These color measurements can then be assembled into a spectrum of the exoplanet’s atmosphere. This unique method does not rely on the use of a reference star which eliminates all achromatic systematic, non-common path errors, including atmospheric amplitude scintillation (the dominant noise source in bright star photometry). The elimination of these noise sources improves the photometric accuracy by up to $20\times$ and produces photometric precisions on-par with space based measurements. This allows for precise, repeatable ($\sim200ppm$) measurements of exoplanet atmospheres from modest ground based observatories. A complete description of the CMI technique, a theoretical framework and its capabilities is described in Limbach et al. in prep\cite{Limbach2021}. 

\section{The Design of ETSI}\label{designETSI}
In this section we discuss the design of the ETSI instrument. We begin with a overview of the instrument's design parameters. We then discuss ETSI's optical layout in section \ref{opDesign}, followed by a detailed discussion of prism and multi-band filters (section \ref{FilterDesign}) design. Finally we briefly discuss the ETSI dectector systems in section \ref{detect}.

We begin with an overview of the ETSI instrument. ETSI is a new type of instrument designed specifically for taking transmission spectra of transiting exoplanets. The Exoplanet Transmission Spectroscopy Imager (ETSI) instrument uses a prism and multi-band filter to simultaneously image 15 spectral bandpasses from 430-975nm during exoplanet transits. The average spectral resolution of the instrument is $R = \lambda/\Delta\lambda = 23$ (the resolution varies for each band with a min $R = 13$ and max $R = 44$). The median bandwidth is 26nm, with the shortest band on the blue end ($\Delta\lambda = 10nm$) and the widest band on the red end ($\Delta\lambda = 75nm$). The bands are generally broader at longer wavelength so that when combined with the lower dispersion with increased wavelength the spot size of each bandpass remains relatively constant. The instrument uses two different sCMOS detectors (discussed more in section \ref{detect}). The fields of view for the detectors are 5.0 and 7.5 square arcminutes so that reference stars can be used (if needed) to detrend long timescale variations in observed stellar flux due to atmospheric airmass and transparency variations. If used, the reference stars are first time-averaged over many exposures ($\sim$ 10 minutes of data) to avoid introducing systematic noise. Table 1 lists the basic ETSI instrument design parameters.

\begin{center}
Table 1: ETSI Instrument Parameters
\begin{center}
\begin{tabular}{ c | c | c}
 ETSI Parameter & Channel 1 Value$^1$ & Channel 2 Value$^1$\\
 \hline
Wavelength Coverage & \multicolumn{2}{c}{430-975 nm}  \\ 
Average Resolution &  \multicolumn{2}{c}{$\lambda/\Delta\lambda = 23$}  \\ 
No. Spectral Measurements & 7 & 8 \\
 Field of View  & $5.0' \times 5.0'$ & $7.5' \times 7.5'$\\
 Plate Scale & 130 mas/pixel & 220 mas/pixel \\
Photometric Accuracy &  \multicolumn{2}{c}{}\\
($5\sigma, 30min, V = 7.5$) & \multicolumn{2}{c}{$<200ppm^2$}
\end{tabular}
\end{center}
\end{center}
$^1${\footnotesize Channel 1 (reflected light from multi-band filter) and channel 2 (transmitted light from octo-chroic) of the instrument use different detectors. This is discussed more in section \ref{detect}.}\\
$^2${\footnotesize Based on prototype on-sky testing\cite{Limbach2021}, Photometric accuracy are expected to improve with ETSI relative the prototype.}

\subsection{Optical Design}\label{opDesign}

\begin{figure}
\centering
\includegraphics[width=0.95\textwidth]{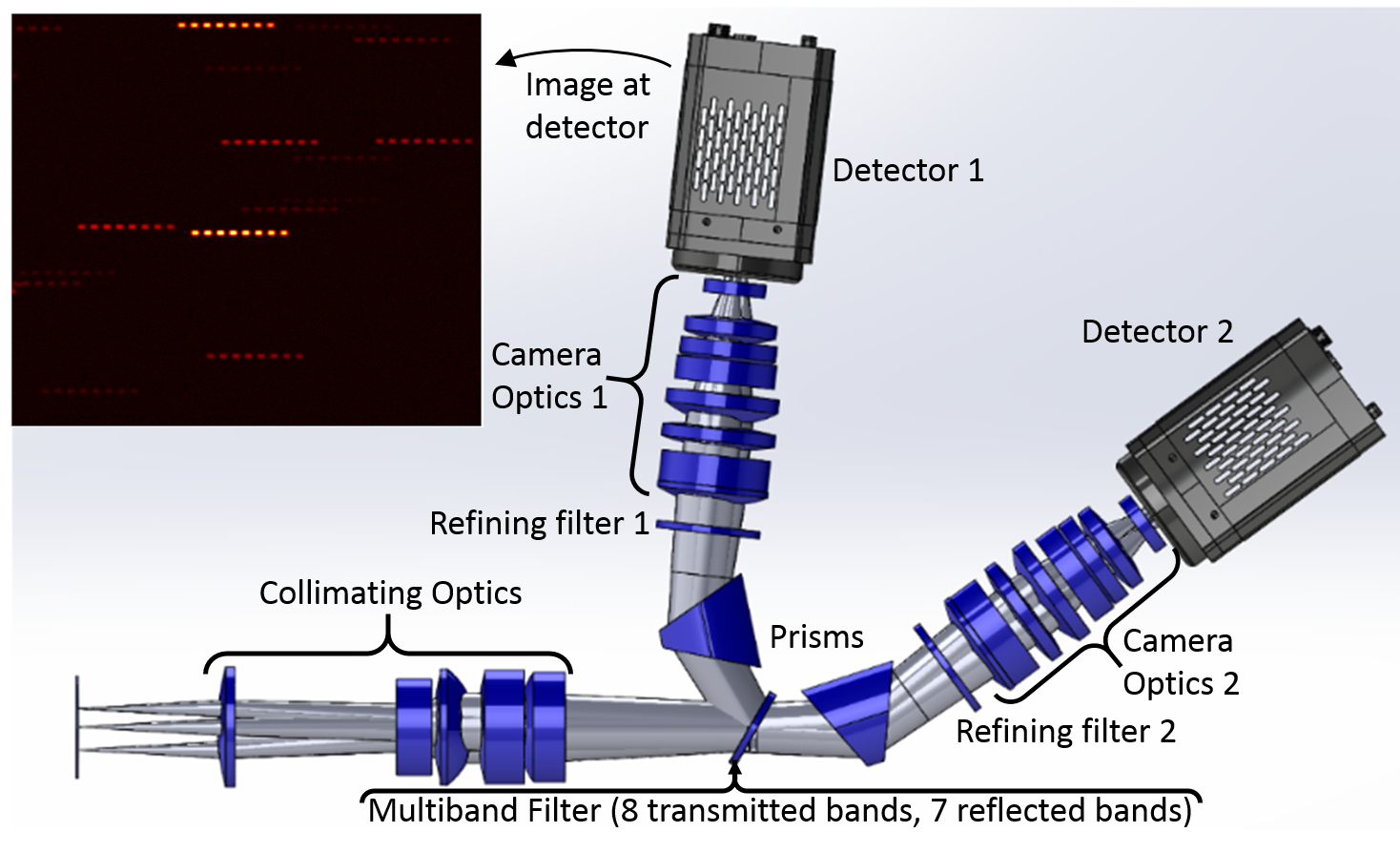}
\caption{The Optical Layout the ETSI instrument. The multi-band filter splits seven and eight spectral bands in to channels one and two of the instrument, respectively. The spectral bands are separated and further cleaned up by a second refining filter. An illustration of a 8-band image using the CMI method is shown in the upper left. Each line of 8 spectral measurements corresponds to one star in the field of view.}
\label{opticalLayout}
\end{figure}

A optical layout of the instrument is shown in Figure \ref{opticalLayout}. Light enters the instrument at the focal plane shown on the left side of the image. A set of collimating lenses then collimate the beam. A novel picket fence multi-band interference filter (discussed more in section \ref{FilterDesign}) then separates out 15 spectral bands (reflecting 7 bands to channel one of the instrument and transmitting 8 bands to the other, alternating every other band). A prism then separates all spectral bands by dispersing them into a line. The light then passes through an additional multi-band refining/clean-up filter to improve the out of band blocking. A set of camera lenses then focus the light onto the detectors producing 7 or 8 images of each star on the image, each from a unique spectral band. The optical layout of a ETSI is shown in Figure \ref{opticalLayout} including a simulated example on-sky 8-band image. A detailed discussion of how this imagery is processed and the resulting photometric precision of exoplanet transit spectra measurements is given in Limbach et al. in prep.\cite{Limbach2021}. In the upper left-hand corner of Figure \ref{opticalLayout} is a simulated image at the detector that images the 8 spectral bands (the bands colored gray in Figures \ref{1000k} and \ref{600k}). Each line of PSFs (or chopped spectrum) corresponds to a different star in the image, and each PSF is a different spectral band. This simulator does not yet include the real dispersion profile of the ETSI prism, so the spot spacing and size will vary from what is shown here in the final instrument.

One unique aspect of CMI is that it requires relatively loose tolerances due do insensitivity to systematic, non-common path errors. Optically, the instrument is relatively insensitive to wavefront error. The specifications for the optics are $<\lambda/4$ irregularity and a scratch/dig of 20/10 (a tighter spec on the scratch/dig is necessary to minimize scattered light which adds to the background noise). Opto-mechanically, the requirements for decenter and lens spacing are both $\pm100 \mu m$, and the tip/tilt requirement for mounting the lenses is $\pm0.5^\circ$. Once mounted, there are a few stability requirements that need to be held to avoid introducing variable chromatic errors (although the precisions required are still fairly minimal). This includes holding the tip/tilt stability of ETSI to $\pm0.1^\circ$ to avoid large variations in the angle of incidence on the interference filter which would introduce time-variable filter bandpass shifts. This would in turn introduce errors in the measured exoplanet spectrum due to spectral band shift induced flux variations from the star. This issue is mostly alleviated by ensuring that none of the ETSI bandpass cut-on/offs occur near sharp stellar features which would lead to a large change in flux even for a small shift in bandpass. It is also important to actively control the focus of the star to avoid focus changes larger than $\gtrapprox 50\mu m$. This tolerance can be fairly loose: if aperture photometry is used to measure flux, variations in focus do not change the measured flux since the PSFs are isolated. However, in practice, we found that fitting the PSFs using more sophisticated methods produced higher precision results and in this case large changes in focus negatively impact ETSI's photometric precision. Therefore, it is necessary to actively control the focus at some level.

The tolerances for the ETSI optics and opto-mechanics illustrate the simplicity of a CMI instrument. This is in stark contrast to other exoplanet detection and characterization techniques which typically require extremely high stability (RV measurements), exquisite wavefront control (direct imaging) and/or rigorous control of systematic noise (transit measurements) to achieve results with precisions similar to the CMI technique.
 
\subsection{Multi-band Filter and Prism Design}\label{FilterDesign}

\begin{figure}
\centering
\includegraphics[width=0.92\textwidth]{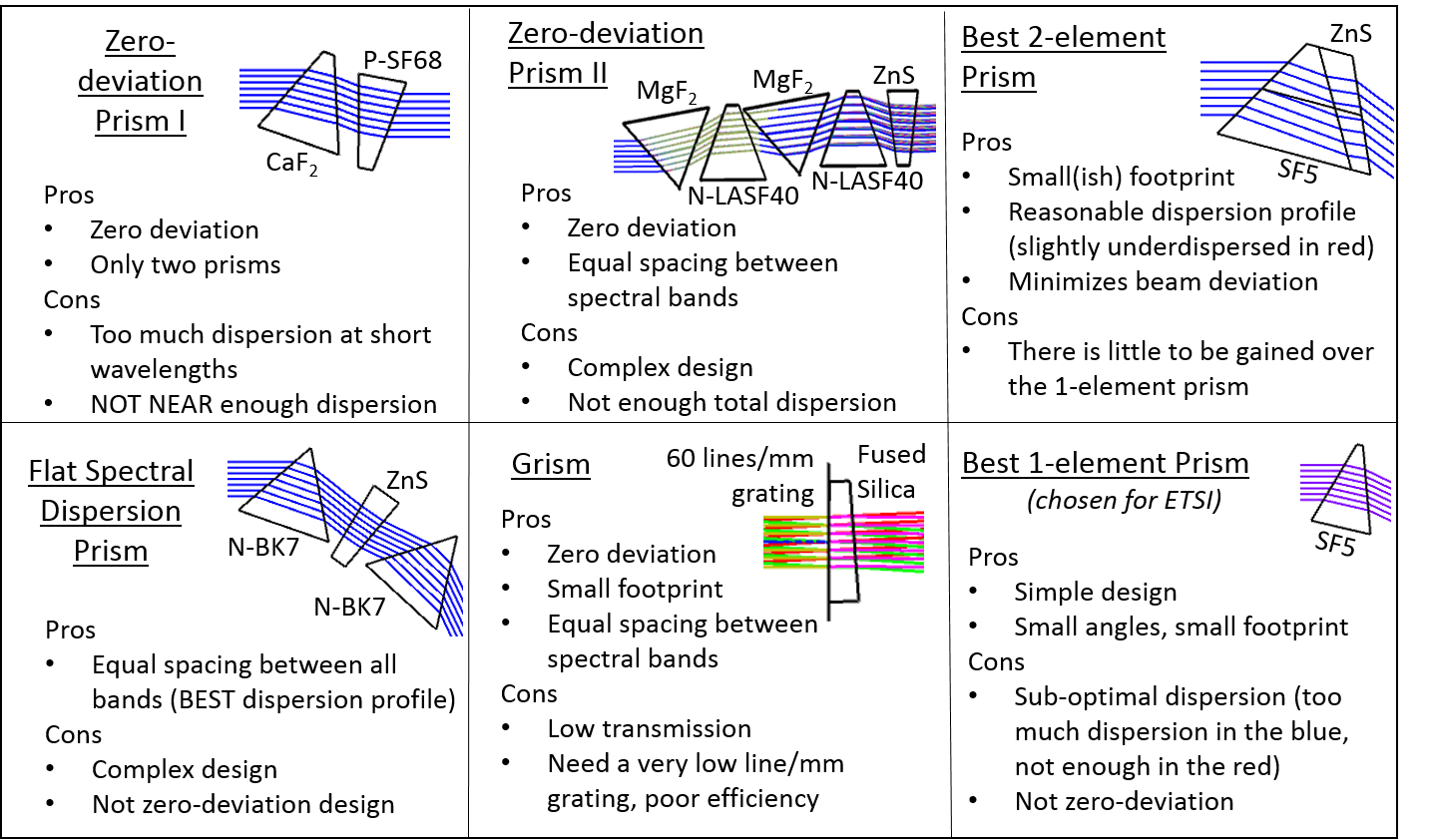}
\caption{This Matrix shows various prism and grism designs we considered for the ETSI instrument along with their pros and cons. The simple 1-element prism on the bottom right was chosen for the ETSI instrument}
\label{prismpics}
\end{figure}

The design process for the filter and prism was iterative. Ultimately we found that the design was driven by the prism because ETSI requires a relatively large amount of dispersion and there are relatively few optical solutions that provide the dispersion required. Ideally, we would have preferred to design a prism set that provided zero-deviation and a flatter dispersion curve across the visible spectrum. For a lower resolution instrument, this would have been possible, but for the relatively large dispersions required for ETSI we found the only practical option was a single element prism with a glass material chosen to provide both a large amount of total dispersion and maximize the amount of dispersion at the longer wavelengths. We also briefly looked at using a grism for the dispersion element. However, this is not a good option for ETSI because  the large ETSI wavelength coverage would cause the grating orders to overlap, the line spacing of the grism would be very low (at the limit of what is currently manufacturable), and the grism would have comparatively low transmission which would negatively impact our photometric precision. Although the grism was not a good choice for ETSI, it may be the right option for higher spectral resolution CMI instruments. Figure \ref{prismpics} shows the layout of several prisms we considered for the ETSI instrument and a brief descriptions of their attributes. From this figure it is immediately apparent that any attempt to use multiple prisms to achieve zero deviation or a flat spectral dispersion curve leads to a very complex prism design and still sub-optimal performance. Although the dispersion profile of the single prism is not perfect (it provides too much dispersion at blue wavelengths) and it requires dealing with a beam deviation in the optical design, given the amount of dispersion required for ETSI, it is the best choice. We found that SF5 glass was optimal for flattening out the dispersion curve across all ETSI wavelengths.

\begin{figure}
\centering
\includegraphics[width=0.91\textwidth]{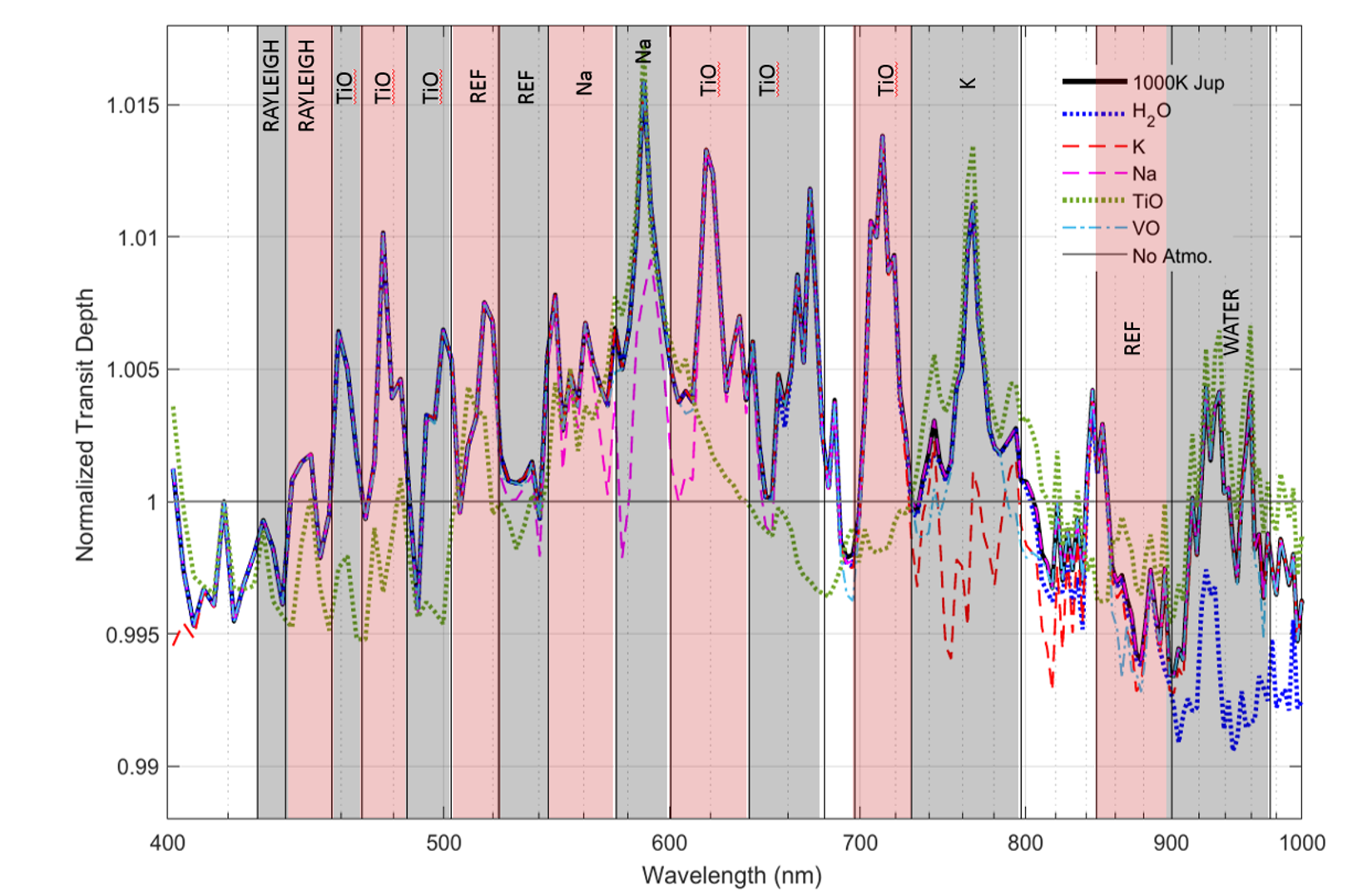}
\caption{Spectrum of a hot Jupiter (T = 1000K), black line, simulated with ExoTransmit\cite{Kempton2016}. Other colored lines are the same model with various species (see legend) removed to show their spectral contribution. The fifteen ETSI spectral bands are overlaid. The 8 gray bands are transmitted to detector 2 and the 7 pink bands are reflected to detector 1. Labels at the top of the plot indicate the species that might be detectable with ETSI in each spectral band when observing a hot jupiter.}
\label{1000k}
\end{figure}

\begin{figure}
\centering
\includegraphics[width=0.91\textwidth]{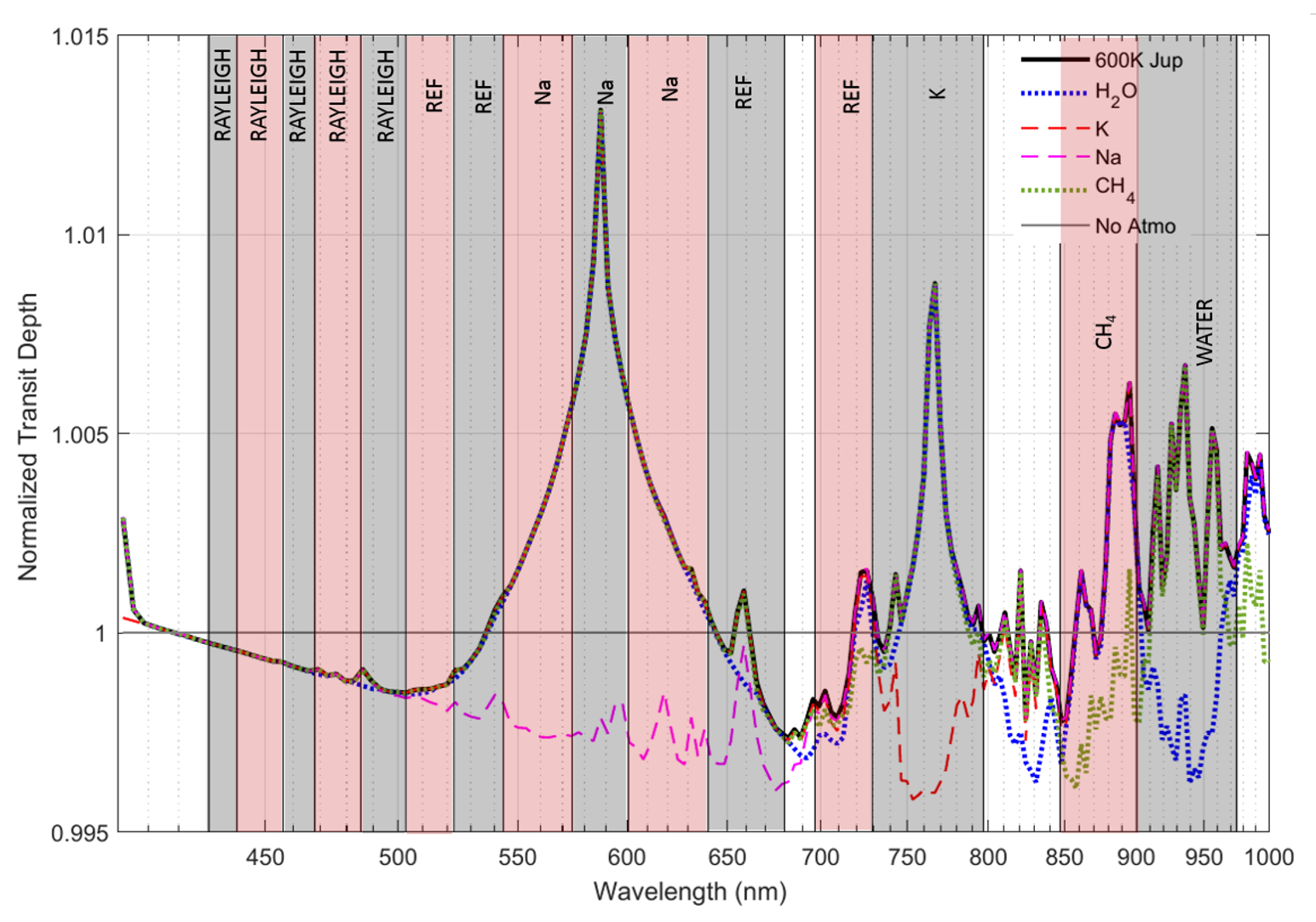}
\caption{Spectrum of a Jupiter-sized exoplanet with T = 600K, black line, simulated with ExoTransmit\cite{Kempton2016}. Other colored lines are the same model with various species (see legend) removed to show their spectral contribution. The fifteen ETSI spectral bands are overlaid. The 8 gray bands are transmitted to detector 2 and the 7 pink bands are reflected to detector 1. Labels at the top of the plot indicate the species that might be detectable with ETSI in each spectral band when observing a T = 600K jupiter.}
\label{600k}
\end{figure}

Once the prism was chosen, the next task was to identify features in exoplanet atmospheres that could be resolved by the spectral dispersion of the ETSI prism and isolated by the multi-band filter. Using Exotransmit\cite{Kempton2016} code, we identified 15 exoplanet spectra bands of interest (shown in Figures \ref{1000k} and \ref{600k}) and designed a 15-band filter for ETSI to image these bands. The filters we have designed for ETSI will allow us to measure Rayleigh scattering, sodium, potassium, methane, water, TiO and clouds in the atmospheres of exoplanets. Hot Jupiters typically provide high SNR data and are the easiest targets for exoplanet transmission spectroscopy. We ensured that our filter captured many of the peaks and troughs expected in a spectrum of a hot Jupiter (Figure \ref{1000k}). However, ETSI should also be able to take spectra of cooler Jupiter sized exoplanets, so we were also careful to ensure that we are sensitive to the unique features in cooler (such as the T=600K Jupiter in Figure \ref{600k}) features such as the methane band between 850-900nm. We were also careful to avoid placing the cut-on/off of our spectral bands near sharp stellar features as this could cause erroneous variations in flux do to instrument flexure. 

To provide a gap between each spectral band, which is the key to ETSI’s sensitivity, a multi-band filter splits every other band to a different channel of the instrument. Seven of the bands (red bars in Figures \ref{1000k} and \ref{600k}) will be imaged to one detector and eight of the bands (gray bars in Figures \ref{1000k} and \ref{600k}) will be imaged to the second detector. Alluxa Inc. is currently manufacturing the ETSI filters - a complex process which is described in more detail below. The selected ETSI prism is placed after the filter in each channel of the instrument to separate each spectral band. Since the prisms disperses more at shorter wavelengths, the shorter wavelength bands are closer together (and the redder bands further apart) so that all the spot sizes are roughly the same in each waveband at the image plane. For large dwarf stars (roughly G stars and larger) this leads to a roughly similar amount of flux in each band since G-dwarf’s have a peak flux near the bluer end of the ETSI wavelength coverage. For M- and K-dwarfs there is a smaller amount of flux in the blue bands, which may necessitate combining data from a couple of the blue spectral bands to achieve significantly high SNR for these stars.

\subsubsection{Design \& Manufacture of the ETSI Filter}\label{FiltMan}
%by Mike Scobey

A critical optical component in the ETSI instrument is the multi-band filter. The design and manufacturing of such a filter is at the cutting edge of thin film technology. In this section, we briefly discuss this process.

Thin film optical interference filters are known to provide the sharpest spectral profiles and lowest loss of any spectral discrimination technology. They are composed of thin stacks of alternating high and low index dielectric materials, each on the order of a quarter wavelength of light in optical thickness. An optical filter design uses constructive and destructive interference in the thin film stacks such that selected bands of wavelengths are either transmitted or rejected via reflection or absorption. In general, the more layers the steeper the slopes and higher the blocking in the out of band regions, although transmission tends to drop, and passbands acquire higher amplitudes of spectral ripple as thickness increases. Computer aided techniques are typically used to create designs. Alluxa further adjusts its designs to enable manufacturing using its proprietary thickness control systems. Filters are generally deposited using Physical Vapor Deposition (PVD) in vacuum chambers and layers are controlled to within a few nanometers of physical thickness. 
 
\begin{figure}
\centering
\includegraphics[width=0.8\textwidth]{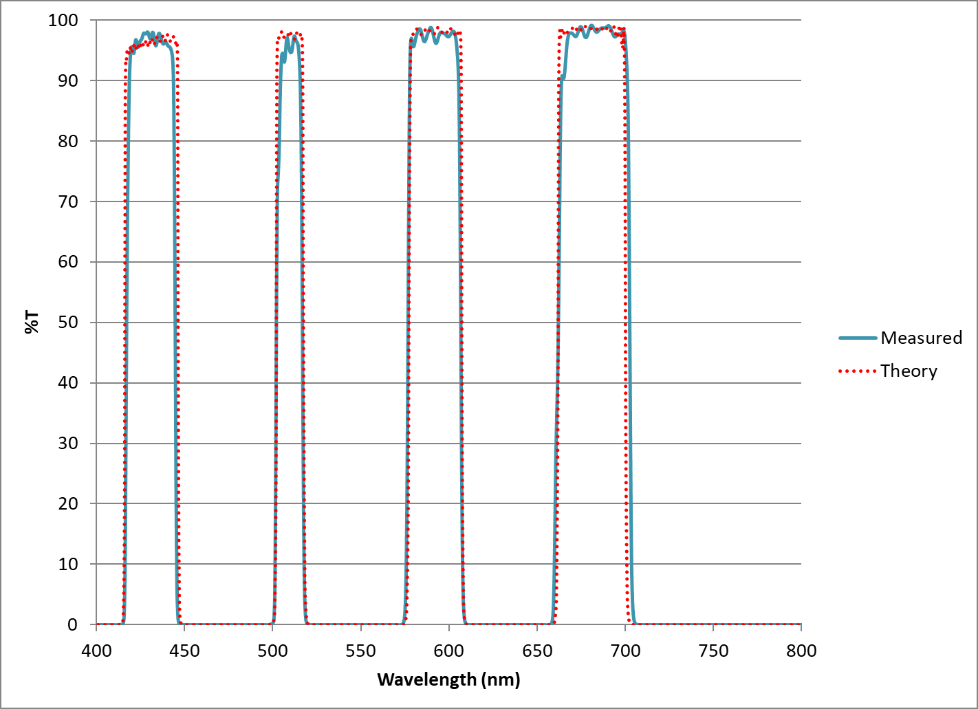}
\caption{Alluxa Quad-band filter, theory and measured, for initial TAMU exoplanet measurements.}
\label{QuadBandFilter}
\end{figure}

Recent advances in optical coating technology have enabled multi-band filter sets of four and five bands. They are now becoming more common for use in applications such as fluorescence detection of multiple marker dyes in life sciences applications. The initial testing of the ETSI used one such filter set, a four band, or “quad-band” filter from Alluxa’s ULTRA series, part number 432-30\textunderscore509.5-16\textunderscore592-30\textunderscore681-38 OD6 ULTRA. This filter provides very high transmission of approximately 95\% in band and average blocking between bands of optical density of 6 (OD6) as shown in Figure \ref{QuadBandFilter}. Although this filter had relatively wide blocking bands and pass bands, the deep blocking required layer counts of 365 layers applied to both sides of a fused silica substrate. The Sirrus plasma PVD technology was used to deposit the filters which create fully dense, humidity shift free, and virtually lossless films, and is capable of several thousand layers on a filter substrate. Tantalum pentoxide is used for the high index and silicon dioxide for the low.

The new 8 band filter set for ETSI will also use the Sirrus PVD coating technology with the same coating materials, and will be custom designed to transmit eight bands and reflect interleaved seven bands between the passbands. The set of filters starts with an ultra-flat beamsplitter, that we are calling an “octo-chroic” (filter 1 of the set) which separates the eight narrow bands in a transmitted beam from a reflected beam composed of the seven interleaved bands. Each beam is further filtered down stream with an additional clean up (or refining) filter used in transmission to provide a total of 3 OD or more blocking between the bands and OD4 out to 1200 nm.  
 
Flatness of the octochroic is maintained at less than 0.1 waves per inch RMS to provide minimum wave front error on the reflected beam. This is achieved through low stress coating processes combined with careful annealing. A broad band high performance anti-reflection (AR) coating with average reflection of less than 0.25\% is deposited on the backside to reduce ghost reflections. Temperature induced variation in wavelength is expected to be less than 10 ppm inherent in the use of hard dielectric coating materials. The fused silica substrates of all three filters are polished to a provide a transmitted wave front error of less than 0.1 waves per inch RMS.
 
Filter 2 is designed to clean up the reflection from the octochroic and is composed of a total of 556 layers. Filter 3, designed to clean up the transmitted band, is slightly simpler and is composed of 232 layers. The total measured performance of the as-built filters in the two beam paths is shown in Figure \ref{reflect} and Figure \ref{transmit} below. Transmitted wavefront error (TWE) on these is less than 0.1 waves per inch RMS.

\begin{figure}
\centering
\includegraphics[width=0.85\textwidth]{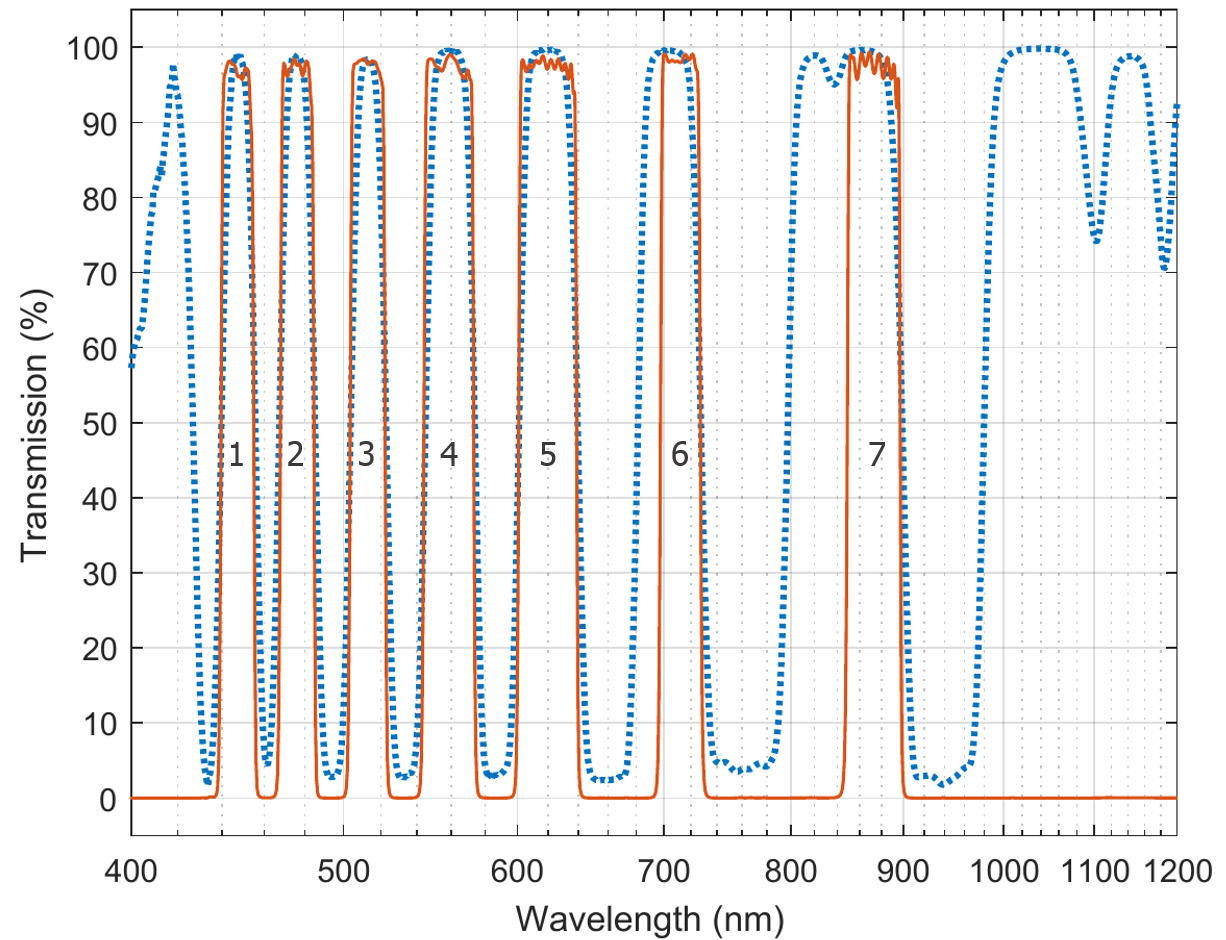}
\caption{Measured transmission of filter 2 path, reflection of Octo-chroic (blue dotted line) and transmission of Filter 2 (red solid line).  Seven bands imaged at the reflected channel detector are labeled by numbers.}
\label{reflect}
\end{figure}

\begin{figure}
\centering
\includegraphics[width=0.85\textwidth]{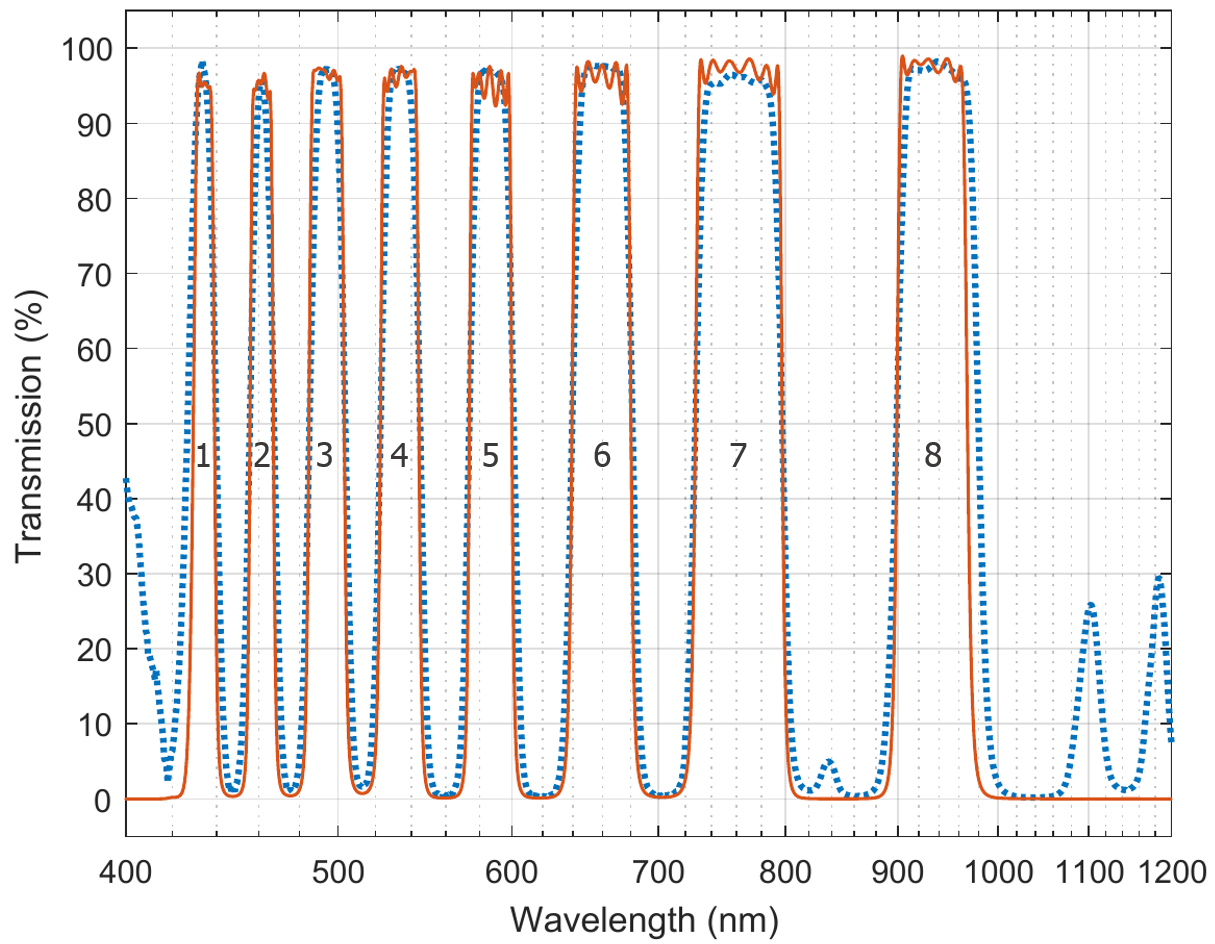}
\caption{Measured transmission of filter 3 path, transmission of octo-chroic (blue dotted line) and transmission of Filter 3 (red solid line). Eight bands imaged at the transmitted channel detector are labeled by numbers.}
\label{transmit}
\end{figure}

\subsection{Detector Trade}\label{detect}
ETSI will use two different sCMOS detectors with one in each channel of the instrument. The sCMOS detectors were chosen for ETSI because their fast readout times allow for an increased total exposure time. For some of our brightest (5th mag) stars on a 2 meter telescope, exposures can be as short as 100ms. If a CCD was used for ETSI, most of the transit observation time would be lost reading out the detector, which would lead to less data and lower photometric precisions. Further, even for faint targets (10th mag), we only need 10 second exposures, which is still well within the exposure time capability of sCMOS cameras. Scientific CMOS detectors are still fairly new to the astronomical community. The use of two different sCMOS detectors will help us determine which is optimal for the ETSI instrument as well as the optimal detector characteristics for our observations. We have spent a considerable amount of time investigating the optimal detector system for ETSI, and intend to select the optimal sCMOS detector system for exoplanet transmission spectroscopy.

\section{Science with ETSI}\label{science}
%by Cole/Chelsea/MA
Once built, the plan is to use ETSI to conduct an exoplanet atmospheres transit survey at McDonald Observatory on a 2 meter telescope beginning in mid-2021. ETSI will observe dozens of exoplanet transits to characterize their atmospheres. In this section, we describe the ETSI survey observing plan and targets (section 4.1) and discuss the expected results from the survey (section 4.2).

\subsection{Observing Plan}
%Chelsea and Cole START EDITING here! Please fill in all the XX's and tables and make plots. Please edit text if needed (especially if I described your code incorrectly. Cole - I used some of the text you wrote up already, please edit if I made any transcription errors.
To develop an observing plan, we need to identify which transiting exoplanets can be characterized with a high signal-to-noise ratio ($SNR > 20\sigma$). Using Tapir\cite{Jensen2013}, we identify targets with transits that are observable from McDonald observatory. We currently constrain our search to our first year of expected ETSI observations (roughly April 2021 - March 2022). For these constraints, we extract the transit information for all objects (including those listed in the NASA exoplanet database as well as TESS target of interests or TOIs). 

Once we have compiled a database of all observable transit during this time-frame, we calculate the expected ETSI spectrum SNR for each target. This code uses information about the total observable transit time during the one year, the exoplanet transit depth and the stellar magnitude. %Elaborate or add more code input description/parameters here if needed
Although we have developed code to accurately estimated ETSI's performance based on many more input parameters (this code is discussed in the next section), this purpose of this code is only to get a rough estimate of ETSI's SNR. Therefore, we only use these three input parameters. Further, we assume the flux from the star is constant in each band. Here we assume an average ETSI value (i.e. the mean instrument transmission, and the mean flux across a 20 nm band from a K-dwarf star). Also, for this calculation we assume we are limited only by photon noise, which is a good assumption for stars brighter than $\sim 13th$ magnitude\cite{Limbach2021}. This calculation is done by assuming (based on more detailed modeling discussed in the next section) that the flux, F, in photons per second from the target star measured by our detector is
\begin{equation*}
    F = \Big(3\times10^7\Big)\Big(10^{2(4.77-m)/5}\Big) photons/s
\end{equation*}
where $m$ is the magnitude of the target star. Then the signal was multiplied by the transit depth, $T$. Further, for this simple model, we assume that only $1\%$ of the transit signal is due to the planet's atmosphere, so we multiply the remaining signal by 0.01 to obtain the number of photons that are useful for characterizing the exoplanet's atmosphere. Since ETSI is limited by the shot noise of the detector the noise in the SNR is simply the square root of the photons per second received from the target star. This gives the signal-to noise-ratio as
\begin{equation*}
    SNR \approx \frac{0.01tFT}{\sqrt{tF}} = 0.01T\sqrt{tF}
\end{equation*}
where $t$ is the exposure time in seconds.

\begin{figure} 
\centering
\includegraphics[width=0.85\textwidth]{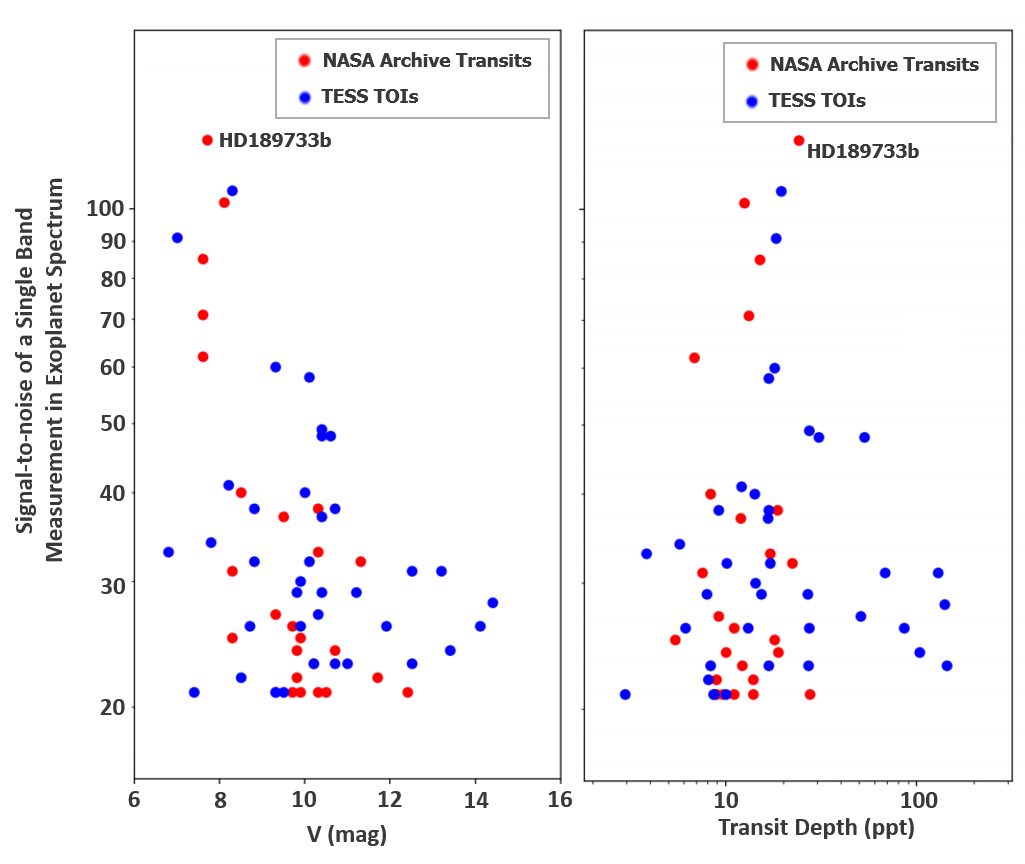}
\caption{Targets observable by ETSI at McDonald observatory. SNR verses Stellar magnitude (left) and SNR verses transit depth (right). Calculated SNR assumes all transits that are observable in one year  are measured by ETSI and contribute to the precision of the planet's spectrum. The given SNR is the average SNR across all 15 bands. Only targets with SNR $>20$ are plotted here.}
\label{ObsPlanets}
\end{figure}

This code produces a list of targets with $SNR>20$ that we can observe with ETSI. For the April 2021 - March 2022 time-frame, there are 25 objects that meet this criteria in the NASA exoplanet database as well as 38 TESS TOIs. Figure \ref{ObsPlanets} shows the achievable ETSI SNR as a function of stellar magnitude (left) and transit depth (right). Each dot in this figure represents an exoplanet (or TESS object of interest). The dots are color-coded with confirmed transiting exoplanets from the NASA archive as red dots and TESS TOIs as blue dots. All of these objects are roughly Jupiter-sized ($R = 0.9 R_{jup}$ or larger). Almost all (92\%) targets are brighter than V = 13 mag, indicating that ETSI's photometric precision will almost always be limited by photon noise. Further, 50\% of these targets are brighter than V = 10 mag, a region that is typically limited in precision by atmospheric scintillation, but will remain photon limited due the the unique CMI method used in the ETSI instrument\cite{Limbach2021}. In mid-2021, ETSI will begin observation of these 63 objects beginning with the highest SNR targets. 

ETSI is a new type of instrument that relies on the common-path multi-band imaging (CMI) technique\cite{Limbach2021}, so initially, ETSI will focus observations on the objects with the highest expected SNR to demonstrate the instrument is performing as expected and is capable of producing reliable and reproducible high SNR exoplanet transmission spectra. Specifically, HD189733b is predicted to have a very high SNR and has been well characterized by many transit spectra measurements and reproducing these results will be ETSI's first task. The top ten targets with the highest SNRs (which will likely be the first objects observed with ETSI) are listed in Table 2.

The targets listed in Table 2 are all Jupiter-sized exoplanets orbiting main sequence stars. The observations of just these ten planets should produce results similar to Sing et al. 2016\cite{Sing2016} (but at a slightly lower resolution and different wavelength coverage). This will provide a new template of what "typical" jupiter atmospheres look like, which will help inform theory/modeling. Further, this will provide the first measurement of many TESS TOIs. The spectra can be used to both confirm if the TOI is indeed an exoplanet as well as characterize the exoplanet's atmosphere. 

The ETSI survey should last approximately two years. The instrument should go on to measure dozens of exoplanet spectra (beyond just the ten described above) on the McDonald 2 m telescope. If successful, the ETSI results will be used to produce the first large database of exoplanet transmission spectra, which would have a profound impact on moving the exoplanet spectra and characterization field forward.

%please fill in this table
\begin{center}
Table 2: The list of Top 10 Exoplanets and TESS TOIs with the Highest ETSI SNRs$^1$
\begin{tabular}{ c | c | c | c | c | c | c | c | c}
   &  & Transit &  & & Est.  & No. Obs.  & Transit & Transits\\
  & $V_*$ & Depth & $R_{planet}$ & $T_{planet}$ %Cite sources for radius and temperature
  & ETSI &  Transits & Duration & obs. for\\
Object  &  (mag) &  (ppt)%or whatever unit you want
& $(R_{Jup})$ & (K) & SNR$^2$ & in 1 yr &  (hrs) &  SNR $>20^3$\\
 \hline
HD 189733 b & 7.7 & 24.0 & 1.13 & 1100 & 125 & 26 & 1.83 & 1\\ %I'm guessing this will be on the list
TIC 328350926.01 & 8.3 & 19.4 & 1.94 & 2500 & 106 & 62 & 1.35 & 2\\ 
WASP-33 b & 8.1 & 12.5 & 1.59 & 2800 & 102 & 59 & 2.85 &  2\\
TIC 20182165.01 & 7.0 & 18.2 & 1.48 & 1000 & 91 & 9 & 3.27 & 1\\
HD 209458 b & 7.6 & 15.0 & 1.39 & 1100 & 85 & 18 & 3.02 & 1\\
KELT-20 b & 7.6 & 13.1 & 1.74 & 2300 & 71 & 16 & 3.58 & 1\\
KELT-9 b & 7.6 & 6.8 & 1.89 & 4100 & 62 & 41 & 3.92 & 3\\
TIC 184679932.01 & 9.3 & 17.9 & 1.61 & 1700 & 60 & 33 & 2.75 & 3\\
TIC 240613164.01 & 10.1 & 16.8 & $-$ & 5100 & 58 & 49 & 4.13 & 5\\
TIC 70914192.01 & 10.4 & 27.3 & 1.76 & 2500 & 49 & 60 & 1.02 & 9\\
\end{tabular}
\end{center}
$^1${\footnotesize The parameters for stellar magnitude, transit depth, planet radius and temperature in this table are based on data from the NASA Exoplanet Archive, which is operated by the California Institute of Technology, under contract with the National Aeronautics and Space Administration under the Exoplanet Exploration Program. Exceptions are the given temperatures for HD 189733b\cite{Knutson2007} and HD 209458b \cite{Deming2005}.}\\
$^2${\footnotesize Estimated ETSI SNR is calculated based on all observable transits from McDonald Observatory in one year.}\\
$^3${\footnotesize Number of complete transits that must be observed to obtain a spectrum with $SNR>20$.}\\
%Chelsea and Cole STOP EDITING here!

%MA to edit below once above section is complete:

\subsection{Expected Science Results}

We have developed an ETSI simulator that takes into account the details of our instrument, it's environment and the target we are observing, and with this information produces a simulated ETSI spectrum. The simulator uses two separate pieces of code to produce these simulations.

\begin{figure}
\centering
\includegraphics[width=0.89\textwidth]{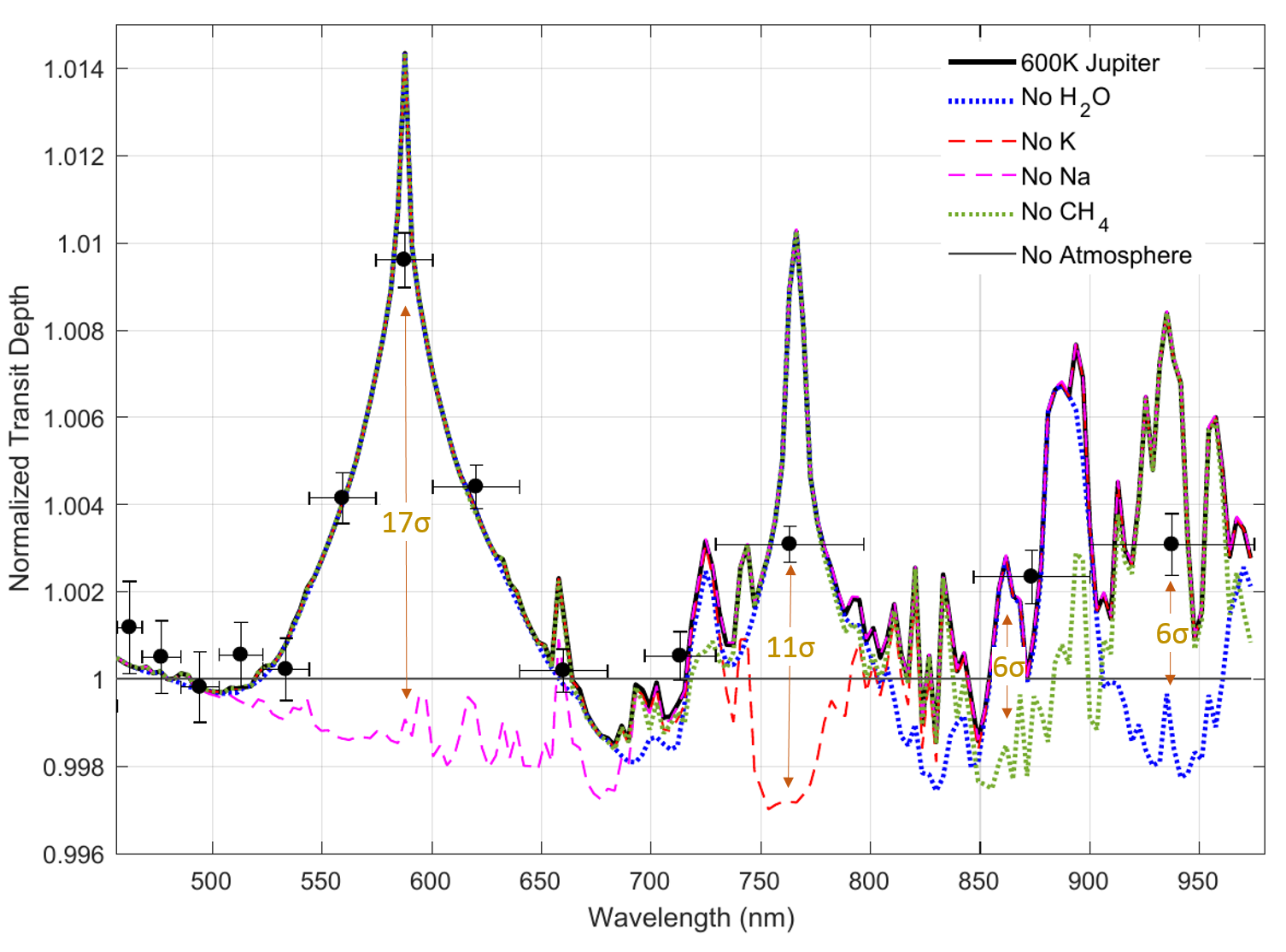}
\caption{Simulated ETSI spectrum (black dots with error bars)  of a R = Rj, T = 600K planet orbiting a V = 9 mag sun-like star. Lines on plot correspond to different atmospheric models: a nominal model (black line) and several models with various species removed to illustrate their spectral imprint. Simulation assumes 2 hours of observations on a 2.7 m telescope during the exoplanet's transit. This figure illustrates that the four molecules/atoms ($H_2O$, $CH_4$,$K$ and $Na$) examined in this simulation are detectable with a SNR $>5\sigma$ in two hours. }
 \label{600kSIM}
\end{figure}

The first piece of code calculates SNR for a given exoplanet. This includes calculating the expected flux to be measured in each of the ETSI bands. The code takes into account the telescope size, the transmission of the atmosphere, telescope and instrument as well as the quantum efficiency of the ETSI detector. For a given target it then calculates the expected signal based on stellar temperature, size and distance to target (the code assumes blackbody stellar radiation) and includes transmission losses. The noise is calculated based on sky background and photon noise. With the calculated signal and noise, the code then uses a given exposure time and (average) transit depth, to calculate the expected SNR of the target in each spectral band of the ETSI instrument. This SNR is used as an input to the second piece of code. We have also used this piece of code to characterize the parameter space that the ETSI instrument is sensitive to.

The second piece of code uses the SNR of each spectral band and a simulated spectrum for the exoplanet of interest to produce synthetic spectra. The simulated spectra are produced from Kempton et al. 2016\cite{Kempton2016}. This simulation code allows us to turn on and off several molecules and atomic species in the exoplanets atmosphere to determine their detectability. It also allows us to vary atmospheric features such as Rayleigh scattering and cloud deck height to determine ETSI's sensitivity to these variables. To create a synthetic spectrum, we produce many spectrum for the exoplanet of interest varying the parameters above. Then, using our calculated SNR per spectral band, we simulated what an ETSI spectrum would look like for the nominal case. We are able to then compare ETSI's performance in each band to the nominal case verses varied models to determine ETSI's sensitivity to specific species and atmospheric features.

Figure \ref{600kSIM} shows the product of our simulator code for a T = 600K, Jupiter-like planet orbiting a V = 9 mag sun-like star. The illustrates that, for this example case, ETSI would be capable of detecting four atoms/molecules ($H_2O$, $CH_4$,$K$ and $Na$) with an SNR $>5\sigma$. For this simulation, we assumed two hours of observations on a 2.7 m telescope.

ETSI remains near-photon limited photometric performance even for bright stars where atmospheric amplitude scintillation noise (and other systematics) typically limit precision. This allows for exquisite spectral measurements of exoplanet atmospheres in a short amount of time on a small telescope, as is simulated here. This makes ETSI optimal for exoplanet transit spectroscopy, especially follow-up on bright targets such as those being identified from the TESS mission. 

\section{Conclusion}
In this paper, we discussed the Exoplanet Transmission Spectroscopy Imager (ETSI) which makes use of a new technique, common-path multi-band imaging (CMI). We examined the optical design of the ETSI instrument. This included the detailed design of the prism and novel multi-band filter which enable ETSI to obtain unprecedented photometric precision during transit spectroscopy measurements. We discussed the expected science with ETSI including our observing plan and simulated ETSI exoplanet transit spectra. We illustrated that dozens of exoplanets transit spectra can be measured with the ETSI instrument on a small (2.7m) telescope in two years. This instrument has the ability to revolutionize the exoplanet transmission spectroscopy field by more than doubling the number of measured exoplanet transmission spectra and producing uniform catalog of gas giant exoplanet spectra. This instrument will also test the new CMI technique. If successful, future instruments similar to ETSI can be used to survey smaller (earth and neptune sized) exoplanets from larger observing facilities. ETSI and the CMI technique may offer a gateway for characterizing almost all transiting exoplanets and potential habitability from ground-based observatories.

\section*{Acknowledgments}
The ETSI instrument is funded by NSF MRI Award Number 1920312. Texas A\&M University thanks Charles R. '62 and Judith G. Munnerlyn, George P. '40 and Cynthia Woods Mitchell, and their families for support of astronomical instrumentation activities in the Department of Physics and Astronomy.

{}

\end{document}